\documentstyle[sprocl]{article}

\def\Journal#1#2#3#4{{#1} {\bf #2}, #3 (#4)}

\def\JHEP{\rm J. High Energy Phys.}

\def\NPB{{\rm Nucl. Phys.} B}
\def\PLB{{\rm Phys. Lett.}  B}

\def\PRD{{\rm Phys. Rev.} D}

\def\npb#1#2#3{\Journal{\NPB}{#1}{#3}{#2}}
\def\prd#1#2#3{\Journal{\PRD}{#1}{#3}{#2}}

\def\plb#1#2#3{\Journal{\PLB}{#1}{#3}{#2}}
\def\jhep#1#2#3{\Journal{\JHEP}{#1}{#3}{#2}}

\def\hth{hep-th}
\def\Hth#1#2#3#4#5#6#7{}

\def\be{\begin{equation}}
\def\ee{\end{equation}}
\def\bea{\begin{eqnarray}}
\def\eea{\end{eqnarray}}

\newcommand{\vev}[1]{\langle#1\rangle}
\def\be{\begin{equation}}
\newcommand{\bel}[1]{\begin{equation}\label{#1}}
\def\ee{\end{equation}}

\newcommand{\rem}[1]{}

\def\NN{{\cal N}}

\def\none{$\NN=1$}

\def\nfour{$\NN=4$}
\def\susy{supersymmetry}
\def\susic{supersymmetric}

\def\rarr{\rightarrow}

\def\ZZ{{\bf Z}}

\begin{document}

\begin{minipage}[t]{3in}
\begin{flushleft}
IASSNS--HEP--98/73\\
hep-ph/9808073 \\
\end{flushleft}
\end{minipage}

\bigskip

\bigskip

\title{ON PHASES OF GAUGE THEORIES AND THE ROLE OF NON-BPS SOLITONS IN
FIELD THEORY}

\author{MATTHEW J.~STRASSLER}

\address{School of Natural Sciences\\Institute for Advanced Study \\ 
Olden Lane\\
Princeton, NJ 08540, USA\\E-mail: strasslr@ias.edu}


\maketitle\abstracts{As shown in hep-th/9709081,\cite{spinmono}
non-BPS saturated solitons play an important role in the duality
transformations of ${\cal N}=1$ supersymmetric gauge theories.  In
particular, a massive spinor in an $SO(N)$ gauge theory with massless
matter in the vector representation appears in the dual description as
a magnetic monopole with a $\ZZ_2$ charge.  This claim is supported by
numerous tests, including detailed matching of flavor quantum numbers.
This fact makes it possible to test the phase of an $SO(N)$ gauge
theory using massive spinors as a probe.  It is thereby shown
explicitly that the free magnetic phase which appears in
supersymmetric theories is a non-confining phase.  A fully non-abelian
version of the Dual Meissner effect is also exhibited, in which the
monopoles are confined by non-BPS string solitons with $\ZZ_2$
charges.\footnote{Talk given at the third workshop on "Continuous
Advances in QCD'', University of Minnesota, Minneapolis, Minnesota,
April 16-19, 1998.}}


After the work of Seiberg \cite{seibone,NAD} and of Intriligator and
Seiberg,\cite{kinsone,kinstwo} we now have strong evidence that \none\
\susic\ gauge theories exhibit a number of different phases at zero
temperature.  A theory may be in the ``Non-Abelian Coulomb Phase''
(NACP), in which case it becomes an interacting conformal field theory
(CFT) in the infrared.  (A given CFT is often the low-energy limit of
two or more gauge theories; these theories are said to be ``dual'' or,
equivalently in this context, ``in the same universality class''.)
Alternatively, a gauge theory may be in the ``Free Magnetic Phase,''
in which case it flows to strong coupling and is best described in
terms of dual variables; these variables are composite quarks and
gluons making up a different, infrared free gauge theory.  There are
confining phases with and without chiral symmetry breaking; in both
cases the infrared physics is again best described using dual
variables, which are composite scalars and fermions made from gauge
singlet combinations of the original fields.  Or a theory may be in
the Higgs phase, with expectation values of charged scalars breaking
its gauge symmetries.  If the theory is not asympotically free, then
it will become weakly coupled in the infrared; this is called the
``Free Electric Phase.''  There are others, but this list will suffice
for present purposes.

In non-\susic\ gauge theories, the situation is far less clear.
Nature provides us, in $SU(3)\times SU(2)\times U(1)$, with examples
of confinement with chiral symmetry breaking, the Higgs phenomenon,
and an infrared-free gauge theory.  We also know that the NACP appears
at large $N_c$ for $N_f = {11\over 2} N_c(1-\epsilon)$, in which case
the beta function can be shown in perturbation theory to have a fixed
point at two loops.

But what about the free magnetic phase, or a confining phase without
chiral symmetry breaking?  Can these also appear in non-\susy\ gauge
theories?  Unfortunately, few analytic tools are at our disposal.
This leaves lattice gauge theory as our only near-term option, where
these questions are very hard to study.  Perhaps at next year's QCD
conference...

We might also ask whether non-\susic\ theories have duality.  The answer,
in some sense, is already known to be ``yes''.  The relation between QCD
and the Chiral Lagrangian --- between quarks and gluons of the short
distance theory and their long-distance hadronic bound states --- is
nearly identical in form to Seiberg's description\cite{seibone} of
the chiral-symmetry-breaking confining phase in \none\ \susic\ gauge
theories, where the SQCD theory is dual to a theory of Goldstone
bosons and their superpartners.

This is the background in which I would like to address a very limited
question associated with the free magnetic phase.  There has been some
confusion in the literature over the issue of whether the free magnetic
phase is confining.  Specifically, in this phase, the low-energy degrees
of freedom are composite quarks and gluons, which form an weakly interacting
dual gauge theory. What about the original degrees of freedom?  Are
they confined, or if not, how do they behave?

We already have some data on this question. In \none\ $SO(N)$ gauge
theories with matter in the vector (${\bf N}$-dimensional)
representation, the confining phases and Higgs phases are distinct.
The energy between two static sources in a spinor representation of
$SO(N)$ (which cannot be screened by massless fields in the vector or
adjoint representation) grows linearly in one case and falls off
exponentially in the other.  What happens in the free magnetic phase?
If the dual gauge theory is {\it abelian}, then one can show that the spinor
of the original theory will appear in the $SO(2)$ dual as a
magnetically charged monopole (which in this case will carry an
additive charge, in contrast to the monopoles we will see later.)
Monopoles in unbroken $SO(2)$ gauge theories are not confined.  This
led to the conjecture that the free magnetic phase is not a confining
phase even when the dual theory is non-abelian.\cite{nsewone,kinsrev}

However, there have been some counterarguments.  It has been noted
that in the free magnetic phase, there are massless mesons, bilinear
in the original fields, which are weakly coupled propagating particles
with a canonical kinetic term.  Does this not suggest confinement?
Furthermore, the confining and free magnetic phase have many
similarities. There are long-range forces in both phases, due to the
massless mesons in the confining case and due to mesons and gluons in
the free magnetic case, so this issue does not distinguish them.  From
this point of view, the dual Lagrangian of the confining phase merely
looks like a special case of the free magnetic phase: an infrared-free
description with long-range interactions.

In the remainder of this talk,  I will show that these counterarguments 
are mistaken, by demonstrating directly that the free magnetic phase
does not confine electric degrees of freedom.  In doing so, I will
show how non-BPS solitons play an important role in \none\ duality,
and I will present a fully non-abelian example of the Dual Meissner effect.

The theories whose phases I will study are \none\ \susic\ $SO(N)$
gauge theories with $N_f$ matter fields $Q^i$ in the ${\bf N}$
representation.  Seiberg \cite{NAD} and Intriligator and Seiberg
\cite{kinstwo} discussed the duality properties of these theories.
The dual description involves the gauge group $SO(N_f-N+4)$, with
$N_f$ fields $q_i$ in the ${\bf N_f-N+4}$ representation, along with
gauge singlet fields $M^{ij}$ and the superpotential $W=M^{ij}q_iq_j$.
A condition of duality is that all gauge invariant operators in one
theory appear as gauge invariant operators in the other.  In this
case, the operator $Q^iQ^j \leftrightarrow M^{ij}$, while
$W_\alpha^nQ^{N-2n} \leftrightarrow \tilde W_\alpha^{2-n}q^{N_f-N+2n}$
(here $W_\alpha$ and $\tilde W_\alpha$ are the field strength
superfields of the electric and magnetic gauge theories).  The
operator $q_iq_j$ is equal, by the equations of motion for $M^{ij}$,
to a total derivative; it is redundant (zero in the infrared) and need
not be mapped.\cite{NAD}

These theories exhibit a number of different phases, of which I will
only focus on two.  If ${3\over 2}(N-2)\geq N_f\geq N-2$ the theory is
in the free magnetic phase; the infrared theory has gauge group
$SO(N_f-N+4)$.  Note that for $N_f=N-2$ the magnetic theory is
abelian.  For $N_f=N-3,N-4$ the theory is confining and has
a vacuum with no chiral symmetry breaking.

To probe the low-energy physics of the theory, I would like to
introduce a pair of static sources and study the energy as a function
of their separation.  The use of sources in the adjoint or vector
representations is not productive, since both will be screened by the
massless particles of the theory.  However, spinor sources will not be
screened.  In short, we should study the Wilson loop in the spinor
representation.  That is fine, so far as the electric theory goes; but
into what is this Wilson loop mapped by duality?  

To deduce the answer to this question, I will use the following trick.
I will first add to the theory a massless dynamical field in the
spinor representation.  The duality in this case is
known.\cite{ppspin,ppmsone} Then, giving the spinor a very large mass,
so that I can treat it as a static source, I will determine how it
appears in the magnetic theory.

To do this, I must outline the relevant duality transformation.\cite{ppmsone}
Consider $SO(8)$ with $N_f$ fields $Q^i$ in the ${\bf 8}$
representation and one field P in the ${\bf 8'}$ (a ``spinor'', if we
call the $Q^i$ ``vectors'' of $SO(8)$.)\footnote{Properly speaking,
the group is $spin(8)$.  The distinction between $SO(8)$ and $spin(8)$
is important, and is properly treated in REF.  In this talk I have
glossed over this issue.}  To this we associate the dual theory
$SU(N_f-4)$ with the following fields: $s$ in the symmetric tensor
representation, $N_f$ fields $q_i$ in the antifundamental
representation, and gauge singlets $M^{ij}$ and $U$.  The
dual superpotential is $M^{ij}q_isq_j + U\det s$.  The operators $M^{ij}$
and $U$ are the images under duality of $Q^iQ^j$ and $P^2$.

How is this duality consistent with the previous one?  Let us add to
the electric theory a mass for the spinor: $W_{elec} \rightarrow
mP^2$.  The effect on the magnetic theory is that $W_{mag} \rightarrow
M^{ij}q_iq_j + U(\det s+m)$.  The equations for a supersymmetric
vacuum include $\partial W/\partial U=0$, so this now implies that
$\vev{\det s}\neq 0$.  Analysis of the D-term potential shows that the only
solution is $\vev{s} = {\bf 1}$.  This expectation value breaks
$SU(N_f-4)$ to $SO(N_f-4)$, leaving the fields $q_i$ as vectors and
the fields $M^{ij}$ as singlets under the gauge group.  The fields $s$
and $U$ become massive.  Thus, the massless fields and superpotential
are precisely those of a theory which is Seiberg's dual of $SO(8)$
with $N_f$ vectors.

But where is the massive spinor particle in the dual theory?  The
breaking of $SU(N_f-4)$ to $SO(N_f-4)$ has a non-trivial topological
property, because the mappings of the two-sphere into $SU(k)/SO(k)$,
$k>2$, form two inequivalent homotopy classes --- {\it i.e.},
$\Pi_2[SU(k)/SO(k)] = \ZZ_2$.  This means that there can be a monopole
soliton, carrying a $\ZZ_2$ magnetic charge, in the dual theory.\cite{LRW,exactmono}
(Recall that in the 't Hooft--Polyakov monopole, we have the breaking
pattern $SU(2)\rightarrow SO(2)$; in this case, since
$\Pi_2[SU(2)/SO(2)] = \ZZ$, the monopole carries an integer magnetic
charge.)  Note that this monopole is not BPS saturated, both because
it carries a non-additive charge and because there are no BPS bounds
for particles in \none\ \susic\ theories.

It is tempting to identify this monopole with the spinor particle of
the original $SO(8)$ theory, and it is possible to provide
considerable evidence for such a conjecture.

First, the spinor and monopole masses are correlated.  In the absence
of BPS bounds, neither mass can be computed in the full quantum
theory.  Semiclassical calculations give the bare spinor mass in the
original theory as $m$ at weak coupling, while that of the monopole in
the dual theory is of order $\vev{s}\sim m^{1/(N_f-4)}$.  We learn
from this that both masses increase with $m$, although, as is to be
expected, quantum corrections to the spinor mass are large.  As $m$
goes to zero, the spinor becomes light, the $SU$ gauge group is
unbroken, and the monopole is lost.  As $m$ goes to infinity, the $SU$
gauge group is broken at ultra-high energies, and the mass of the
monopole correspondingly goes to infinity (and its size to zero.)

A second piece of evidence regards the $\ZZ_2$ charges of the spinor and
monopole.  The monopole has $\ZZ_2$ charge as a result of topology.  The
spinor has $\ZZ_2$ charge in the following sense: although it carries
quantum numbers in the spinor representation of $SO(8)$, most of these
quantum numbers are screened by the light fields in the vector
representation and in the adjoint representation.  Only a global $\ZZ_2$
quantum number --- spinor number --- will be unscreened.  This $\ZZ_2$
is the subgroup of the $\ZZ_2\times \ZZ_2$ center of $SO(8)$ under
which all the massless fields are neutral.  Similar relations, connecting
topological charges of monopoles with charges of electric
states under the center of the dual gauge group, underly Olive-Montonen
duality in \nfour\ \susic\ gauge theories.\cite{GNO,Osborn,rdew,mskyo}
Here we learn that they play a role in \none\ duality as well.

A third and powerful check involves matching flavor quantum numbers of
both spinors and monopoles.  Suppose we consider $SO(8)$ with a
massive spinor, and allow $k$ fields in the vector representation to
acquire expectation values: $\vev{Q^1Q^1} = \cdots =
\vev{Q^kQ^k}$. This breaks $SO(8)\times SU(N_f)$ to $SO(8-k)\times
SO(k)\times SU(N_f-k)$, where the $SO(8)$ and $SO(8-k)$ are gauge
groups and the other group factors are global flavor symmetries.  The
${\bf 8'}$ spinor representation becomes a spinor under both the gauge
group $SO(8-k)$ and the flavor group $SO(k)$; for example, if $k=3$ it
becomes a $({\bf 4,2})$ of $SO(5)\times SO(3)$.  While the gauge
quantum numbers of the spinor should not be carried by the monopole, as
only gauge invariant states have meaning under duality, flavor quantum
numbers should be visible in both the original and in the dual theory.
In particular, if the spinor transforms under a flavor symmetry, then the
monopole must do so as well.

Miraculously, this check works perfectly, and in a way which
intricately depends on the details of the duality.  Recall that the
dual theory has gauge group $SU(N_f-4)$ broken to $SO(N_f-4)$ and
superpotential $W= M^{ij}q_isq_j+ U(\det s+m)$.  The expectation
values for the operators $Q^iQ^i$ lead to trilinear terms
$\vev{M^{ij}} q_isq_i$, $i=1,\dots, k$.  This breaks the global
symmetry under which the $q_i$ and $M^{ij}$ transform from $SU(N_f)$
to $SO(k)\times SU(N_f-k)$, as in the original theory.  
The monopole, which is built by winding the expectation value of the
field $s$ around infinity, acquires {\it fermionic zero modes} in the
presence of trilinear $q_isq_i$ couplings, one for each of the $k$
fields which have such a term.  These zero modes cause the monopole to
transform in the spinor representation of $SO(k)$, in agreement with the flavor
transformations of the field $P$ in the original theory.

A couple of other checks are worthy of mention.  On physical grounds,
if the theory has several fields in the spinor representation, one
should only see $\ZZ_2$ monopoles in the theory if {\it all} the
spinors are massive.  This is because a massive spinor will be
screened if there are massless spinors in the theory, and thus there
will be no physical states carrying $\ZZ_2$ charge.  This constraint
is satisfied: the breaking pattern $SU(k)\rarr SO(k)$ only occurs when
all of the spinors are massive, and so monopoles only appear when they
are expected.  In addition, although I have discussed only $SO(8)$, the
spinor/monopole relation and the associated consistency checks
generalize to $SO(10)$.\cite{ppmstwo,mltispnr,spinmono}

Having established the relation beyond reasonable doubt,
\footnote{Actually this statement is tongue-in-cheek.  There are a
number of additional physical subtleties which need to be discussed,
such as the question of whether the monopole is the lightest state
with magnetic charge, before one may be fully confident that this
relation has been properly interpreted.  These issues are fully
addressed in Ref. 1, to which the interested and/or skeptical
reader is referred.}  I now wish to apply this result to the study of
phases in $SO(8)$ gauge theories.  For this purpose we need static
sources in the spinor representation.  We may obtain such sources by
taking the mass of $P$ to infinity.  In the dual theory this involves
taking the mass of the monopole to infinity and its size to zero,
turning it into a static pointlike Dirac monopole with a $\ZZ_2$
charge.  In other words, the Wilson line in the spinor representation of
$SO(8)$ is mapped under duality to the $\ZZ_2$-valued 't Hooft loop in
the dual $SO(N_f-4)$ description.

It is now trivial to show that the free magnetic phase is not a
confining phase.  For $SO(8)$ with $N_f = 7,8,9$ fields in the vector
representation, the dual $SO(3),SO(4),SO(5)$ gauge theory is infrared
free.  Magnetically charged point sources in an unbroken non-abelian
infrared-free gauge theory are unconfined, just as in abelian gauge
theories.  It follows that the free magnetic phase is not associated
with confinement of electric charge, and that the confining phase is a
completely distinct phase.

Next consider adding masses to some of the vectors, so that only
$N_f=5$ of them remain massless.  This process causes the dual
$SO(N_f-4)$ gauge group to be broken completely.  Because $\pi_1[SO(N)] =
\ZZ_2$,\footnote{That is, there are two classes of maps of the circle
into $SO(N)$, as in the case of $SO(3)$, which is the three-sphere
with north and south poles identified.} the breaking of this group
leads to string solitons (magnetic flux tubes) carrying a $\ZZ_2$
quantum number.  Again, the existence of these string solitons is protected by
topology, but they are not BPS saturated.  The $\ZZ_2$-charged Dirac
monopole sources are confined by these flux tubes.  The original
$SO(8)$ theory thus has electric flux tubes carrying $\ZZ_2$ charge,
which confine the spinor sources.  Note this $\ZZ_2$ charge is the
expected one for confinement of spinors in $SO(8)$ with vectors; it is
the part of the center of $SO(8)$ under which all light fields are
neutral.

We have here a fully non-abelian realization of the Dual Meissner
effect; both the electric (confining) and magnetic (Higgsed) theories
are non-abelian.  Similar examples have been discussed\cite{mskyo}
in the context of breaking pure \nfour\ gauge theory to pure \none;
there, the non-BPS-saturated $\ZZ_N$ strings of pure $SU(N)$ gauge
theory emerge via the breaking of a dual $SU(N)/\ZZ_N$ gauge symmetry.
Note that the implications for the abelian projection approach to
confinement, popular with some lattice gauge theorists, are not
positive.

To summarize, I have demonstrated that the free magnetic phase is not
a confining phase.  Wilson loops will not have an area law, and thus
this phase is completely distinct from the confining phase and is not
a generalization of the latter. In the confining phase, I have shown
you a fully non-abelian example of the Dual Meissner effect,
suggesting that a similar effect in real QCD might not be realizable
via abelian projection techniques.  In order to study these issues, I
first had to show that duality implies that a massive unscreened
particle in one theory shows up as a magnetic monopole, with a $\ZZ_2$
topological charge, in its dual.  A particularly powerful test was
provided by a study of flavor symmetries.  Finally, and perhaps most
importantly in the long run, the monopole and string solitons which
appeared in this talk, and which played an important role in duality
and the associated physics, were stabilized by topological charges;
none of them was BPS saturated.  This gives a clear indication that
topology plays an essential role even in \none\ duality.  We may hope
that this role will be further clarified in the near future.

A final note: as this talk was in preparation, string theorists
(particularly Sen \cite{nonbpsstring}) began to turn their attention
to this very interesting topic.

\section*{Acknowledgments}
I would like to thank the organizers for inviting me to this 
conference.  The work reported here was 
supported in part by National Science Foundation grant
NSF PHY-9513835 and by the W.M.~Keck Foundation.

\end{document}